\documentclass{llncs}

\usepackage{color}
\usepackage{colortbl}
\usepackage{fancyvrb}
\usepackage{listings}

\definecolor{Gray}{gray}{0.85}
\definecolor{LightCyan}{rgb}{0.88,1,1}
\definecolor{codeBlue}{rgb}{0,0,0.5}
\definecolor{codeGreen}{rgb}{0,0.4,0}
\definecolor{codeRed}{rgb}{0,0.2,0.2}
\definecolor{codeOlive}{rgb}{0.5,0.5,0.2}
\definecolor{codeMacro}{rgb}{0.6,0.2,0.2}
\definecolor{codeGray}{rgb}{0.5,0.5,0.5}

\begin{document}

\title{Parallelware tools: An experimental evaluation on POWER systems}
\titlerunning{Parallelware tools for POWER systems}  
%
\author{Manuel Arenaz\inst{1} \and Xavier Martorell\inst{2} }
\authorrunning{Manuel Arenaz et al.} 
%
\tocauthor{Manuel Arenaz, and Xavier Martorell}
\institute{University of A Coruna and Appentra Solutions, Spain, \email{manuel.arenaz@appentra.com}
\and
Computer Architecture Dept., Universitat Polit\'ecnica de Catalunya and Computer Sciences Dept., Barcelona Supercomputing Center, Spain \email{xavier.martorell@bsc.es }}

\lstset{
	language=C,
    tabsize=3,
    frame=single,
	numbers=left,
	numbersep=5pt,
    breaklines=true,
	numberstyle=\tiny\sffamily\color{codeGray},
    escapeinside={(*}{*)},
    basicstyle=\bfseries\scriptsize\color{codeRed},
    morecomment=[l][{\color{codeOlive}}]{\#i},
    morecomment=[l][{\color{codeOlive}}]{\#d},
    morecomment=[l][{\color{codeOlive}}]{\#e},
    morecomment=[l][{\color{codeOlive}}]{\#e},
    morecomment=[l][{\color{codeOlive}}]{\#p},
    identifierstyle=\mdseries\color{black},
    keywordstyle=\bfseries\color{codeBlue},
    commentstyle=\itshape\color{codeGreen}
}

\maketitle              

\begin{abstract}
Static code analysis tools are designed to aid software developers to build better quality software in less time, by detecting defects early in the software development life cycle. Even the most experienced developer regularly introduces coding defects. Identifying, mitigating and resolving defects is an essential part of the software development process, but frequently defects can go undetected. One defect can lead to a minor malfunction or cause serious security and safety issues. This is magnified in the development of the complex parallel software required to exploit modern heterogeneous multicore hardware. Thus, there is an urgent need for new static code analysis tools to help in building better concurrent and parallel software. The paper reports preliminary results about the use of Appentra's Parallelware technology to address this problem from the following three perspectives: finding concurrency issues in the code, discovering new opportunities for parallelization in the code, and generating parallel-equivalent codes that enable tasks to run faster. The paper also presents experimental results using well-known scientific codes and POWER systems.

\end{abstract}

\section{Introduction}
Static code analysis tools are highly specialized to detect one or more defects, typically categorized into similar types of defects. These tools fulfill a group of specific needs of software developers. It is only recently that heterogeneous multicore systems have been adopted in a wide-range of hardware in industrial sectors such as automotive, wireless communication and embedded vision. Therefore it is increasingly important to develop new static code analyses that address the fundamental problem of concurrency, which means that many tasks running at the same time on the same hardware can lead to unpredictable and incorrect behaviour. Identifying and fixing issues related to concurrency and parallelism is one of the most time-consuming and costly aspects of parallel programming. However, static code analysis tools that detect defects related to parallel programming are at a very early stage.

This papers presents an experimental evaluation of Appentra's Parallelware static code analysis tools on POWER systems, which go beyond the state of the art by addressing the problem of concurrency and parallelism from three different perspectives: finding concurrency issues in the code, discovering new opportunities for parallelization in the code, and generating parallel-equivalent code that enables tasks to runs faster. In the rest of the paper, Section~\ref{sec:parallelware} describes the current set of Parallelware tools, namely, the Parallelware development library, Parallelware Analyzer (BETA) and Parallelware Trainer. Next, Section~\ref{sec:experiments} presents early results from the analysis of the SNU NPB Suite~\cite{SNU-NPB:benchs}, a C version of the NAS Parallel Benchmarks~\cite{NPB:benchs}, using POWER systems available at the J\"ulich Supercomputing Centre and at Appentra headquarters. Finally, Section~\ref{sec:conclusions} presents conclusions and future work.

\section{Parallelware Tools}
\label{sec:parallelware}

Appentra is a Deep Tech global company that delivers products based on the Parallelware technology~\cite{ArenazToplas,ArenazAndion}, a unique approach to static code analysis for concurrent and parallel programming. It is based on an engine for the detection of parallel patterns such as forall, scalar reduction, sparse forall and sparse reduction. These patterns are used to detect software issues related to concurrency and parallelism, discover parallelism and generate parallel-equivalent code. The current portfolio of tools based on Parallelware technology is as follows:
\begin{itemize}
    \item {\bf Parallelware developer library}, which  offers the static code analysis capabilities of the Parallelware technology. It provides an Application Program Interface (API) that is the basis of Parallelware Analyzer and Parallelware Trainer, and that is designed to enable the integration in third-party software development tools. It supports the C programming language, the OpenMP 4.5~\cite{OpenMP4.5:spec} and OpenACC 2.5~\cite{OpenACC2.5:spec} directive-based parallel programming interfaces, and the multithreading, offloading and tasking parallel programming paradigms.
    \item {\bf Parallelware Analyzer (BETA)}~\cite{Parallelware} is designed to speed up the development of parallel applications and to enforce best practice in parallel programming for heterogeneous multicore systems. It helps software developers by finding software defects early in the parallelization process and thus increases productivity, maintainability and sustainability. It is available as a set of command-line tools to enable compatibility with Continuous Integration and DevOps platforms.
    \item {\bf Parallelware Trainer}~\cite{Parallelware} is an interactive, real-time code editor that enables scalable, interactive teaching and learning of parallel programming, increasing productivity and retention of learning. It is available for Windows, Linux and MacOS operating systems.
\end{itemize}

\section{Experimental results}
\label{sec:experiments}

This section presents experimental results obtained on POWER systems using Parallelware tools. More specifically, Section~\ref{sec:experiments:pwanalyzer} presents the report generated by Parallelware Analyzer and Section~\ref{sec:experiments:pwtrainer} presents experimental results of codes parallelized using Parallelware Trainer.

\subsection{Report generated using Parallelware Analyzer}
\label{sec:experiments:pwanalyzer}

The report shown in Table~\ref{table:pwanalyzer} was generated by the Parallelware Analyzer tool after analyzing codes written in the C programming language from the SNU NPB Suite~\cite{SNU-NPB:benchs,NPB:benchs} benchmarks (NPB-SER-C and NPB-OMP-C implementations). The structure of the report is as follows: {\em Benchmark}, the software application; {\em Files}, number of source code files; {\em SLOC}, source lines of code calculated by the \emph{sloccount} tool; {\em Time}, runtime of the Parallelware Analyzer tool in milliseconds; {\em Software issues}, number of issues found in the code related to concurrency and parallelism; and {\em Opportunities}, number of loops found in the code that have opportunities for parallelization using multithreading and SIMD paradigms. The last row of the table provides total numbers for all the analyzed benchmarks.

The current tool setup reports five software issues related to concurrency and parallelism: {\em Global}, use of global variables in the body of a function; {\em Scope}, scalar variables not declared in the smallest scope possible in the code; {\em Pure}, pure functions free of side effects not marked by the programmer; {\em Scoping}, variables in an OpenMP parallel region without an explicit data scoping; and {\em Default}, OpenMP parallel region without the \emph{default(none)} clause. More information about each one of them can be found in the Appentra Knowledge website~\cite{Appentra:knowledge}.The tool also reports two types of opportunities for parallelization: {\em Multi}, outer loops that can be parallelized with the multithreading paradigm; and {\em SIMD}, inner loops that can be parallelized with the SIMD paradigm.

Parallelware Analyzer successfully analyzed a total of 192 source files of code, containing 39890 lines of code written in the C programming language, in less than 13 seconds. In terms of software issues related to concurrency and parallelism, the tools detected a total of 296 uses of global variables in the body of functions. There are 2082 declarations of scalars in a scope bigger than necessary. Moreover, a total of 9 pure functions that are free of side effects but not marked as such were found. Finally, 329 variables with an implicit datascoping and 117 OpenMP parallel regions having a default one were detected. In terms of opportunities for parallelization, a total of 312 outer loops and the same number of inner loops can be parallelized using the multithreading and SIMD paradigms respectively.

\begin{table}[t] \centering\setlength\tabcolsep{3pt}
\begin{tabular}{ | l | r r r | r r r r r | r r |  }
\hline
\hline
Benchmark                   & Files & SLOC & Time  & \multicolumn{5}{|c|}{Software issues} & \multicolumn{2}{|c|}{Opportunities} \\
                            &       &      & (ms) & Global & Scope & Pure & Scoping & Default & Multi & SIMD \\
\hline
\hline

NPB3.3-SER-C/BT & 17 & 2608 & 557.97 & 13 & 143 & 0 & 0 & 0 & 24 & 44 \\
NPB3.3-SER-C/CG & 3 & 521 & 143.69 & 3 & 20 & 1 & 0 & 0 & 13 & 10 \\
NPB3.3-SER-C/DC & 11 & 2725 & 430.67 & 10 & 0 & 3 & 0 & 0 & 13 & 0 \\
NPB3.3-SER-C/EP & 2 & 175 & 88.61 & 1 & 0 & 0 & 0 & 0 & 2 & 0 \\
NPB3.3-SER-C/FT & 7 & 625 & 238.77 & 4 & 0 & 1 & 0 & 0 & 0 & 0 \\
NPB3.3-SER-C/IS & 2 & 463 & 69.3 & 4 & 0 & 0 & 0 & 0 & 4 & 0 \\
NPB3.3-SER-C/LU & 19 & 2389 & 739.86 & 15 & 298 & 0 & 0 & 0 & 29 & 59 \\
NPB3.3-SER-C/MG & 3 & 873 & 648.68 & 11 & 2 & 0 & 0 & 0 & 4 & 2 \\
NPB3.3-SER-C/SP & 19 & 2056 & 683.6 & 19 & 381 & 0 & 0 & 0 & 28 & 91 \\
NPB3.3-SER-C/UA & 13 & 5576 & 2181.73 & 53 & 163 & 0 & 0 & 0 & 77 & 69 \\
NPB3.3-SER-C/common & 0 & 296 & 174.19 & 0 & 0 & 0 & 0 & 0 & 0 & 0 \\
NPB3.3-SER-C/config & 0 & 0 & 28.71 & 0 & 0 & 0 & 0 & 0 & 0 & 0 \\
NPB3.3-SER-C/sys & 1 & 759 & 182.71 & 2 & 0 & 0 & 0 & 0 & 0 & 0 \\

\hline
\hline

NPB3.3-SER-C & 97 & 19066 & 6168.48 & 135 & 1007 & 5 & 0 & 0 & 194 & 275 \\

\hline
\hline

NPB3.3-OMP-C/BT & 17 & 2693 & 568.03 & 13 & 144 & 0 & 41 & 9 & 8 & 4 \\
NPB3.3-OMP-C/CG & 3 & 627 & 171.77 & 5 & 20 & 1 & 16 & 5 & 9 & 3 \\
NPB3.3-OMP-C/DC & 11 & 2754 & 425.05 & 10 & 0 & 3 & 0 & 0 & 13 & 0 \\
NPB3.3-OMP-C/EP & 2 & 198 & 92.4 & 1 & 0 & 0 & 4 & 3 & 0 & 0 \\
NPB3.3-OMP-C/FT & 3 & 649 & 163.39 & 12 & 0 & 0 & 10 & 8 & 0 & 0 \\
NPB3.3-OMP-C/IS & 2 & 634 & 88.15 & 6 & 4 & 0 & 7 & 4 & 4 & 0 \\
NPB3.3-OMP-C/LU & 20 & 2542 & 778.31 & 17 & 295 & 0 & 55 & 9 & 5 & 0 \\
NPB3.3-OMP-C/MG & 3 & 923 & 662.07 & 11 & 2 & 0 & 19 & 10 & 4 & 2 \\
NPB3.3-OMP-C/SP & 19 & 2147 & 693.33 & 19 & 381 & 0 & 45 & 13 & 8 & 4 \\
NPB3.3-OMP-C/UA & 14 & 6549 & 2749.03 & 65 & 229 & 0 & 132 & 56 & 67 & 24 \\
NPB3.3-OMP-C/bin & 0 & 0 & 29.25 & 0 & 0 & 0 & 0 & 0 & 0 & 0 \\
NPB3.3-OMP-C/common & 0 & 349 & 178.34 & 0 & 0 & 0 & 0 & 0 & 0 & 0 \\
NPB3.3-OMP-C/config & 0 & 0 & 28.47 & 0 & 0 & 0 & 0 & 0 & 0 & 0 \\
NPB3.3-OMP-C/sys & 1 & 759 & 179.29 & 2 & 0 & 0 & 0 & 0 & 0 & 0 \\

\hline
\hline

NPB3.3-OMP & 95 & 20824 & 6806.88 & 161 & 1075 & 4 & 329 & 117 & 118 & 37 \\

\hline
\hline

Totals  & 192 & 39890 & 12975.36 & 296 & 2082 & 9 & 329 & 117 & 312 & 312 \\
\hline
\hline

\hline \end{tabular}
\caption{Parallelware Analyzer report.}
\label{table:pwanalyzer}
\end{table}

\subsection{Report generated using Parallelware Trainer}
\label{sec:experiments:pwtrainer}

The Parallelware Trainer tool was used to automatically generate several parallel versions of a code that computes the Mandelbrot sets. Four parallel versions of Mandelbrot are considered in this work: {\em Sequential}, serial version (see Listing~\ref{Mandelbrot:multithreading}, ignoring the OpenMP directives); {\em Multithreading}, OpenMP version using multithreading paradigm (see Listing~\ref{Mandelbrot:multithreading}, which contains directives \emph{\#pragma omp parallel for}); {\em Taskwait}, parallel version using OpenMP 3.0 tasking paradigm (see Listing~\ref{Mandelbrot:taskwait}, which contains directives \emph{\#pragma omp task} and \emph{\#pragma omp taskwait}); {\em Taskloop}, parallel version using OpenMP 4.5 tasking paradigm (see Listing~\ref{Mandelbrot:taskloop}, which contains directives \emph{\#pragma omp taskloop}).  It  should be noted that a software engineer with little experience used the tool to generate and test all the parallel versions for correctness and performance in less than one hour.

Experiments were conducted on two POWER systems: a compute node of the {\em Juron} supercomputer at J\"ulich Supercomputing Centre and the {\em Appentra server} available at Appentra's headquarters. In Juron, the hardware setup of each compute node consists on a IBM S822LC system with 2x 10-core SMT8 POWER8NVL CPUs, offering a total of 160 threads. It provides a {\em CentOS Linux 7 (AltArch)} Linux operating system with a GCC 4.8.5 compiler. In Appentra server, the hardware setup consists on a RaptorCS Talos II system equipped with an 8-core SMT4 POWER9 processor, offering a total of 32 threads. It runs a {\em Debian 10 (buster)} Linux with a GCC 8.3.0 compiler. In both systems, GCC compiler flags were used as follows: {\tt -O2} for sequential execution and {\tt -fopenmp -O2} for OpenMP-enabled parallel execution.

Table~\ref{table:results:mandelbrot} shows the runtimes and speedups for a problem size of 20000. Its structure is as follows: {\em Version}, serial or parallel version of the code, one of {\em Sequential}, {\em Multithreading}, {\em Taskwait} and {\em Taskloop}; {\em No. Threads}, number of OpenMP threads; {\em Time}, runtime in seconds, and {\em Speedup}, speedup calculated with respect to the sequential version, for each POWER system. The {\em Taskwait} version is the fastest code both in Juron (maximum speedup is 56 for 160 threads) and Appentra's POWER9 server (maximum speedup above 28 for 32 and 64 threads). The {\em Multithreading} version is also fast, but the speedup is below {\em Taskwait} because the OpenMP code generated by Paralellware Trainer includes the clause \emph{schedule(auto)} which defaults to \emph{schedule(static)}. Note that since the workload of Mandelbrot is not constant, different threads are assigned different workloads. Therefore, \emph{schedule(static)} is not the better choice and should be replaced by \emph{schedule(static,1)} or \emph{schedule(dynamic)}. Finally, note that the {\em Taskloop} version does not scale with the number of threads. This needs to be further investigated as we expected {\em Taskloop} to also decrease the execution time on both systems.

\begin{table}[!t] \centering\setlength\tabcolsep{3pt}
\begin{tabular}{ | l | r | r r | r r | }
\hline
        &  & Juron   &          &     Appentra's server    &         \\
\hline
Version	& No.Threads & Time (secs) & Speedup  & Time (secs) & Speedup \\
\hline
\hline
Sequential     & 4 &  89.50 & 1    & 178.92 & 1    \\
Multithreading & 4 &  32.85 & 2.72 &  37.94 & 4.72 \\
Taskwait       & 4 &  23.30 & 3.84 &  24.38 & 7.34 \\
Taskloop       & 4 & 133.31 & 0.67 &  37.99 & 4.71 \\
\hline
Sequential     & 8 &  89.52 & 1    & 178.92 & 1    \\
Multithreading & 8 &  31.77 & 2.82 &  38.96 & 4.59 \\
Taskwait       & 8 &  17.91 & 4.99 &  21.57 & 8.29 \\
Taskloop       & 8 & 143.22 & 0.63 &  38.82 & 4.61 \\
\hline
Sequential     & 16 & 89.51 &  1    & 178.93 &  1    \\
Multithreading & 16 & 14.42 &  6.21 &  20.96 &  8.54 \\
Taskwait       & 16 &  7.67 & 11.67 &  12.02 & 14.89 \\
Taskloop       & 16 & 86.44 &  1.04 &  20.99 &  8.53 \\
\hline
Sequential     & 32 &  89.51 &  1    & 178.93 &  1    \\
Multithreading & 32 &   8.57 & 10.45 &  10.93 & 16.37 \\
Taskwait       & 32 &   4.97 & 19.01 &   6.31 & 28.36 \\
Taskloop       & 32 &  99.93 &  0.89 &  11.05 & 16.19 \\
\hline
Sequential     & 64 & 89.52 &  1    & 178.92 &  1    \\
Multithreading & 64 &  4.24 & 21.11 &   7.80 & 22.94 \\
Taskwait       & 64 &  2.60 & 34.43 &   6.33 & 28.27 \\
Taskloop       & 64 & 86.45 &  1.04 &   7.70 & 23.24 \\
\hline
Sequential     & 80 & 89.53 &  1    \\
Multithreading & 80 &  3.50 & 25.58 \\
Taskwait       & 80 &  2.34 & 38.26 \\
Taskloop       & 80 & 86.45 &  1.04 \\
\cline{1-4}
Sequential     & 128 & 89.51 &  1    \\
Multithreading & 128 &  2.59 & 34.56 \\
Taskwait       & 128 &  1.64 & 54.58 \\
Taskloop       & 128 & 86.46 &  1.04 \\
\cline{1-4}
Sequential     & 160 & 89.53 &   1   \\
Multithreading & 160 &  2.53 & 35.39 \\
Taskwait       & 160 &  1.60 & 55.96 \\
Taskloop       & 160 & 86.42 &  1.04 \\
\cline{1-4}
\cline{1-4}
\end{tabular}
\caption{Execution times (in seconds) and speedups of Mandelbrot in Juron (2x 10-core SMT8 POWER8 processors) and in Appentra's POWER server (8-core SMT4 POWER9) for problem size of 20000.}
\label{table:results:mandelbrot}
\end{table}

\section{Conclusions and Future Work}
\label{sec:conclusions}
Preliminary results show evidences that Parallelware tools have the potential to help software developers to build better quality parallel code. On the one side, Parallelware Analyzer was used to evaluate the SNU NPB Suite, a C implementation of the NAS Parallel Benchmarks. The static code analysis capabilities of Parallelware technology reported the existence of data scoping issues in the codes as well as the existence of pure functions which were not marked as such to provide additional hints to the compiler. Additionally, the tool also reported the existence of sequential loops that could be parallelized using the multithreading and SIMD paradigms.

On the other side, Parallelware Trainer provides a GUI that facilitates the generation of parallel version of a code, as well as the testing of those version for correctness and performance. In less than one hour, a software engineer with little experience in parallel programming generated several OpenMP-enabled parallel versions of the Mandelbrot algorithm using multithreading and tasking paradigms. Performance tests showed significant speedups on both Juron and Appentra POWER systems.

As future work, we plan to further develop Parallelware tools to support C++ and Fortran, as well as other task-based parallel versions tuned for execution on GPUs and FPGAs. We also plan to extend the number of software issues related to concurrency and parallelism detected by the Parallelware tools and run them on a wider set of scientific and engineering software.

\section*{Acknowledgements}
\label{sec:acks}
This work has been partly funded from the Spanish Ministry of Science and Technology (TIN2015-65316-P), the Departament d'Innovaci\'o, Universitats i Empresa de la Generalitat de Catalunya (MPEXPAR: Models de Programaci\'o i Entorns d'Execuci\'o Parallels, 2014-SGR-1051), and the European Union’s Horizon 2020 research and innovation program throughgrant agreements MAESTRO (801101) and EPEEC (801051). The authors gratefully acknowledge the access to the Juron system at J\"ulich Supercomputing Centre.

%
%
\bibliographystyle{abbrv}
\bibliography{biblio}


\clearpage

\begin{minipage}{0.9\columnwidth}
\begin{lstlisting}[caption={OpenMP-enabled parallel version of Mandelbrot using multithreading paradigm. Parallel code automatically generated by Parallelware Trainer.}, label={Mandelbrot:multithreading}]
int mandelbrot(int max_iter, int height, int width, double **output, double real_min, double real_max, double imag_min,
               double imag_max) {
    double scale_real = (real_max - real_min) / width;
    double scale_imag = (imag_max - imag_min) / height;

    #pragma omp parallel default(none) shared(height, imag_min, max_iter, output, real_min, scale_imag, scale_real, width)
    {
    #pragma omp for schedule(auto)
    for (int row = 0; row < height; row++) {
        for (int col = 0; col < width; col++) {

            double x0 = real_min + col * scale_real;
            double y0 = imag_min + row * scale_imag;

            double y = 0, x = 0;
            int iter = 0;
            while (x * x + y * y < 4 && iter < max_iter) {
                double xtemp = x * x - y * y + x0;
                y = 2 * x * y + y0;
                x = xtemp;
                iter++;
            }
            output[row][col] = iter;
        }
    }
    } // end parallel
    return 0;
}
\end{lstlisting}
\end{minipage}

\begin{minipage}{0.9\columnwidth}
\begin{lstlisting}[caption={OpenMP-enabled parallel version of Mandelbrot using tasking paradigm of OpenMP version 3.0 (task/taskwait). Parallel code automatically generated by Parallelware Trainer.}, label={Mandelbrot:taskwait}]
int mandelbrot(int max_iter, int height, int width, double **output, double real_min, double real_max, double imag_min,
               double imag_max) {
    double scale_real = (real_max - real_min) / width;
    double scale_imag = (imag_max - imag_min) / height;

    #pragma omp parallel default(none) shared(height, imag_min, max_iter, output, real_min, scale_imag, scale_real, width)
    #pragma omp single
    {
    for (int row = 0; row < height; row++) {
    #pragma omp task
    {
        for (int col = 0; col < width; col++) {
            ...
            output[row][col] = iter;
        }
    } // end task
    }
    #pragma omp taskwait
    } // end parallel
    return 0;
}
\end{lstlisting}
\end{minipage}

\begin{minipage}{0.9\columnwidth}
\begin{lstlisting}[caption={OpenMP-enabled parallel version of Mandelbrot using tasking paradigm of OpenMP version 4.5 (taskloop). Parallel code automatically generated by Parallelware Trainer.}, label={Mandelbrot:taskloop}]
int mandelbrot(int max_iter, int height, int width, double **output, double real_min, double real_max, double imag_min,
               double imag_max) {
    double scale_real = (real_max - real_min) / width;
    double scale_imag = (imag_max - imag_min) / height;

    #pragma omp parallel default(none) shared(height, imag_min, max_iter, output, real_min, scale_imag, scale_real, width)
    #pragma omp single
    {
    #pragma omp taskloop
    for (int row = 0; row < height; row++) {
        for (int col = 0; col < width; col++) {
            ...
            output[row][col] = iter;
        }
    }
    } // end parallel
    return 0;
}
\end{lstlisting}
\end{minipage}

\end{document}